\documentclass[aps,pre,twocolumn]{revtex4-2}
\usepackage[T1]{fontenc} 
\usepackage[utf8x]{inputenc}
\usepackage{graphicx}
\usepackage{amsmath,amsfonts,mathtools}
\usepackage{color}

\begin{document}

\title{Genuine multifractality in time series is due to temporal correlations}

\author{Jarosław Kwapień$^1$}
\email{jaroslaw.kwapien@ifj.edu.pl}
\author{Pawel Blasiak$^{1,2}$}
\email{pawel.blasiak@ifj.edu.pl}
\author{Stanisław Drożdż$^{1,3}$}
\email{stanislaw.drozdz@ifj.edu.pl}
\author{Paweł Oświęcimka$^{1,4}$}
\email{pawel.oswiecimka@ifj.edu.pl}

\affiliation{$^1$Complex Systems Theory Department, Institute of Nuclear Physics, Polish Academy of Sciences, ul. Radzikowskiego 152, 31-342 Krak\'ow, Poland}
\affiliation{$^2$Institute for Quantum Studies, Chapman University, Orange, CA 92866, USA}
\affiliation{$^3$Faculty of Computer Science and Telecommunications, Cracow University of Technology, ul. Warszawska 24, 31-155 Kraków, Poland}
\affiliation{$^4$Faculty of Physics, Astronomy and Applied Computer Science, Jagiellonian University, ul. Łojasiewicza 11, 30-348 Kraków, Poland}

\date{\today}

\begin{abstract}
Based on the mathematical arguments formulated within the Multifractal Detrended Fluctuation Analysis (MFDFA) approach it is shown that in the uncorrelated time series from the Gaussian basin of attraction the effects resembling multifractality asymptotically disappear for positive moments when the length of time series increases. A hint is given that this applies to the negative moments as well and extends to the L\'evy stable regime of fluctuations. The related effects are also illustrated and confirmed by numerical simulations. This documents that the genuine multifractality in time series may only result from the long-range temporal correlations and the fatter distribution tails of fluctuations may broaden the width of singularity spectrum only when such correlations are present. The frequently asked question of what makes multifractality in time series - temporal correlations or broad distribution tails - is thus ill posed. In the absence of correlations only the bifractal or monofractal cases are possible. The former corresponds to the L\'evy stable regime of fluctuations while the latter to the ones belonging to the Gaussian basin of attraction in the sense of the Central Limit Theorem.
\end{abstract}

\keywords{Complexity, Time series, Multifractality, Finite size effects, Central Limit Theorem} 

\maketitle

\section{Introduction}

Multifractality~\cite{mandelbrot1982,stanley1988,barabasi1991} is a concept that allows to compactly grasp the most essential aspects of complexity~\cite{kwapien2012} and it therefore pervades all the science. Indeed, its applications cover a broad range of spatial static structures or dynamical phenomena as represented by the time series $x(t)$~\cite{rosas2002,struzik2002,gagnon2006,dimatteo2007,lopes2009,ausloos2012,dutta2013,drozdz2015,drozdz2016,oswiecimka2018,jiang2019,rak2020,watorek2021a}. The most frequently employed related formal quantitative tool is the multifractal spectrum $f(\alpha)$ which reflects the value of fractal dimension of the support in $t$ of the set of singularities of $x(t)$ carrying a particular value of the H\"older exponents $\alpha$ at the point $t$~\cite{halsey1986,muzy1994}. Value of such an exponent reflects the degree of singularity that a fluctuation of $x(t)$ develops at this particular point. The degree of such a singularity is determined by the amplitude of the related fluctuation but also by the amplitudes of fluctuations in its direct neighborhood and such a correspondence cascades through a sequence of scales. The relative ordering of fluctuations, thus the correlations among them, is as important as their absolute values. The related factors are indecomposable and correlations are nonlinear. It is thus seriously surprising that in the scientific literature still a question is being asked as to which factor makes multifractality with correlations and distributions of fluctuations considered as its two independent 'sources'. In majority of such cases the series are randomly shuffled which of course destroys correlations. The finite width of the multifractal spectrum obtained after such a procedure - thus in the absence of temporal correlations - is then often taken as an evidence that in that particular case multifractality is due to the broadness of tails in the distribution of fluctuations. However, the broad $f(\alpha)$ spectra, observed for relatively short temporal scales if the time series are characterised by heavy-tailed probability distribution functions (pdfs), cannot be interpreted as genuine multifractality. They originate from the synergy of a small sample size for short scales and a subtle ``awareness'' or ``memory'' effect related to the fact that the L\'evy-stable pdfs are characterised by bifractal $f(\alpha)$. For time series with the unstable pdfs, for which the central limit theorem applies, the broad singularity spectra disappear for sufficiently long scales if the time series length allows for considering such scales. Some numerical and empirical evidence of this effect can be found in~\cite{kwapien2005,drozdz2009,zhou2012,rak2018,oswiecimka2020,olivares2022}. The present contribution documents this fact using analytical arguments as well as some corresponding numerical illustrations.   

\section{Formal arguments}

The algorithm used most frequently for quantifying multifractality in time series is based on detrending~\cite{peng1994} and is commonly known as Multifractal Detrended Fluctuation Analysis  (MFDFA)~\cite{kantelhardt2002}. Indeed, this algorithm appears the most stable and practical for such a purpose~\cite{oswiecimka2006}. Its most general extension that even allows for quantification of the multifractal cross-correlations~\cite{zhou2008} between two time series in its consistent variant is known as Multifractal Cross-Correlation Analysis (MFCCA)~\cite{oswiecimka2014} and MFDFA is then its special case. 

\subsection{Multifractal detrended cross-correlation analysis (MFCCA)}

Let us thus assume there are two time series consisting of $T$ data points: $U=\{u_i\}^T_{i=1}$ and $W=\{w_i\}^T_{i=1}$. As the first step of the procedure, for a given temporal scale $s$ ($s_{\rm min} \le s \le s_{\rm max}$), the time series is divided into segments of length $s$ by starting from both ends, which brings $M_s=2 \lfloor{T/s} \rfloor$ segments total, where $\lfloor\cdot\rfloor$ denotes floor equal to integer part in this case. Within each segment $\nu$ ($\nu=0,...,M_s-1$) detrended signal profiles $X$ and $Y$ are constructed by integrating the data points and subtracting a local trend represented by a fitted $m$th-degree polynomial $P_{z,\nu}^{(m)}(i)$:
\begin{eqnarray}
X_i (s,\nu) = \sum_{j=1}^i u_{j+\nu s} - P_{X,\nu}^{(m)}(i),\\
Y_i (s,\nu) = \sum_{j=1}^i w_{j+\nu s} - P_{Y,\nu}^{(m)}(i).
\label{eq::signal.profiles}
\end{eqnarray}
Next in each segment a respective covariance between $X$ and $Y$ is calculated:
\begin{equation}
f_{XY}^2(s,\nu) = {1 \over s} \sum_{i=1}^s [X_i(s,\nu) - \Bar{X}(s,\nu)] [Y_i(s,\nu) - \Bar{Y}(s,\nu)],
\label{eq::covariance}
\end{equation}
where $\Bar{X},\Bar{Y}$ denote averaging, and then the so-defined covariances are used to derive a family of fluctuation functions $F_r^{XY}(s)$ of order $r$:
\begin{equation}
F_r^{XY}(s) = \Big\{ {1 \over M_s} \sum_{\nu=0}^{M_s-1} {\rm sign} \left[ f_{XY}^2(s,\nu) \right] |f_{XY}^2(s,\nu)|^{r/2} \Big\}^{1/r},
\label{eq::fluctuation.function}
\end{equation}
where $\textrm{sign}(x)$ denotes sign of $x$. Factoring out the covariance sign is needed to prevent the expression becoming complex and to preserve information stored in $f_{XY}^2$~\cite{oswiecimka2014}. The fluctuation functions are calculated for a range of different temporal scales $s$, with typical $s_{\rm min}$ larger than the longest sequence of constant signal values and $s_{\rm max}$ equal to $T/5$~\cite{oswiecimka2006}.

The fluctuation functions constitute the basic quantities for detecting the fractal properties of time series, because, if this is the case, one obtains their power-law dependence on scale:
\begin{equation}
F_r^{XY}(s) \sim s^{h(r)}.
\label{eq::hurst}
\end{equation}
A family of thus obtained generalized Hurst exponents $h(r)$ can serve as an indicator of the degree of multifractality. When $h(r) = {\rm const}$, the structure is just monofractal. Otherwise it is multifractal and the broadness of $h(r)$-dependence on $r$ reflects the richness of the convolution of various fractal components in forming the resulting multifractal composition. 

\subsection{Gaussian uncorrelated signals}

We consider the signal profiles $X$ and $Y$ that do not possess any temporal correlations, neither auto- nor cross- and they are described by the same probability distribution. First, we perform rigorous calculations of the bivariate fluctuation functions for $X$ and $Y$. In order to hold the degree of complication of the related explicit inspection of their scale $s$ dependence within the reasonable limits, we consider the case of $r > 0$ here. The general case of $r\in \mathbb{R}$ will be considered later. Without loss of generality, let us also assume that $r=2n$ with $n \in \mathbb{N}$ in order to simplify the calculations (since $h(r)$ is monotonous in $r$, all non-integer values of the exponent $r$ provide us with $h(r)$ that is placed between its values for the nearest integers). 

We intend to work out an analytical assessment of $F_{2n}^{XY}(s)$ for $X,Y$. We thus start by rewriting Eq.(\ref{eq::fluctuation.function}) for $r/2=n$, which allows us to remove the factorization into sign and modulus of $f_{XY}^2$:
\begin{widetext}
\begin{align}
\nonumber
&\left[ F_{2n}^{XY}(s) \right]^{2n} = {1 \over M_s} \sum_{\nu=0}^{M_s-1} \left[ {1 \over s} \sum_{i=1}^s X_i(s,\nu)Y_i(s,\nu)\right]^n=\\
&= {1 \over s^n} \sum_{i_1=1}^s \dots \sum_{i_n=1}^s {1 \over M_s} \sum_{\nu=0}^{M_s-1} X_{i_1}(s,\nu) Y_{i_1} (s,\nu) \cdot ... \cdot X_{i_n}(s,\nu) Y_{i_n} (s,\nu) = {1 \over s^n} \sum_{i_1=1}^s \dots \sum_{i_n=1}^s \mathbb{E}(X_{i_1}Y_{i_1}\cdot ... \cdot X_{i_n}Y_{i_n}),
\label{eq::moments.simplified}
\end{align}
\end{widetext}
where the arithmetic average over $M_s$ segments $\nu$ is replaced by the corresponding moments $\mathbb{E}(\cdot)$. Under the no-memory assumption of the signals, the products $X_g Y_g$ and $X_h Y_h$ are statistically independent for any $g \neq h$ (with $1 \le g,h \le s$), which allows for factorization: $\mathbb{E}(\dots X_g^{l_g} Y_g^{l_g} \dots X_h^{l_h} Y_h^{l_h} \dots) = \mathbb{E}(X_g^{l_g} Y_g^{l_g}) \mathbb{E}(X_h^{l_h} Y_h^{l_h}) \mathbb{E}(\dots)$, where $0 \le l_g,l_h \le n$ and $\sum_h l_h = n$.

This factorization and multiple summation in Eq.(\ref{eq::moments.simplified}) allows us to rewrite it by considering a combinatorial problem of partitioning an $n$-element set $\mathcal{S}$ into $k$ subsets of size $\{l_1,...,l_k \}$ with $1 \le l_p \le n$ and $\sum_{p=1}^k l_p = n$, provided each element of $\mathcal{S}$ can assume one of $s$ values:
\begin{widetext}
\begin{align}
\nonumber
{1 \over s^n} \sum_{i_1=1}^s \dots \sum_{i_n=1}^s \mathbb{E}(X_{i_1}Y_{i_1}\cdot ... \cdot X_{i_n}Y_{i_n}) = \\
= {1 \over s^n} \sum_{j_1=0}^n \dots \sum_{j_n=0}^n & {n! \over (1!)^{j_1} \cdot ... \cdot (n!)^{j_n}} {1 \over j_1! \cdot ... \cdot j_n!} \prod_{l=0}^{k-1} (s-l) \prod_{m=1}^n \mathbb{E}^{j_m} (X^m Y^m),
\label{eq::factorization.simplified}
\end{align}
\end{widetext}
where $\sum_m j_m = k$ and $\sum_m m j_m = n$ (with $0 \le j_m \le n$). Here it is implicitly assumed that the products $X_h Y_h$ have the same probability distribution function for any $h$, which brought $\mathbb{E}(X_g^m Y_g^m) = \mathbb{E}(X_h^m Y_h^m)$ for any $g,h,m$. This, in turn, allowed us to neglect the subscripts of $X,Y$ and led to the product $\prod_l (s-l)$ that is the number of possible $n$-permutations of $s$ possible $g,h$ values.

The r.h.s. of Eq.(\ref{eq::factorization.simplified}) comes from a fact that a partition of $\mathcal{S}$ into $k$ subsets of size $\{l_1,...,l_k \}$ can be realized in $S(l_1,...,l_k)$ ways, where
\begin{widetext}
\begin{align}
\nonumber
S(l_1,...,l_k) & = {n \choose l_k} {n - l_k \choose l_{k-1}} \cdot ... \cdot {n - \sum_{p=2}^{k} l_p \choose l_1} = {n! \over l_k! (n-l_k)!} {(n-l_k)! \over l_{k-1}! (n-l_k-l_{k-1})!} \cdot ... \cdot {(n - \sum_{p=2}^k l_p)! \over l_1! (n - \sum_{p=1}^k l_p)!} =\\
& = {n! \over l_k! l_{k-1}! \cdot ... \cdot l_1!} = {n! \over (1!)^{j_1} \cdot ... \cdot (n!)^{j_n}},
\label{eq::number.ways.simplified}
\end{align}
\end{widetext}
and writing the last equality in Eq.(\ref{eq::number.ways.simplified}) required a reorganization only ($0 \le j_p \le n$). Moreover, since the subsets with the same number of elements should not be distinguishable, we have to divide the final form of $S(l_1,...,l_k)$ by $j_1! \cdot ... \cdot j_n!$, which completes the explanation of Eq.(\ref{eq::factorization.simplified}).

Now we may consider different probability distribution functions describing the random variables $X,Y$. A special case is the 
bivariate Gaussian p.d.f., for which we assume $\mathbb{E}(X)=\mathbb{E}(Y)=0$ and put $\mathbb{E}(XY)=\sigma_{XY}^2$, $\mathbb{E}(X^2)=\sigma_X^2$, $\mathbb{E}(Y^2)=\sigma_Y^2$. By using the Isserlis theorem~\cite{isserlis1916} generalized to higher-order moments (the well known Wick's probability theorem~\cite{wick1950} is its related variant), we write:
\begin{widetext}
\begin{align}
\mathbb{E}(X^m Y^m) = \sum_{p=0}^{\lfloor m/2 \rfloor} {(m!)^2 \over 2^{2p} (m-2p)! (2!)^p} \sigma_X^{2p} \sigma_Y^{2p} \sigma_{XY}^{2(m-2p)} = \sum_{p=0}^{\lfloor m/2 \rfloor} {(m!)^2 \over 2^{3p} (m-2p)!} \sigma_X^{2p} \sigma_Y^{2p} \sigma_{XY}^{2(m-2p)},
\label{eq::isserlis.simplified}
\end{align}
\end{widetext}
where $\lfloor \cdot \rfloor$ denotes integer part. After inserting Eq.(\ref{eq::factorization.simplified}) and Eq.(\ref{eq::isserlis.simplified}) into Eq.(\ref{eq::moments.simplified}), we obtain a complete form of the $r$-fluctuation function $F_r^{XY}(s)$:
\begin{widetext}
\begin{align}
\left[ F_{2n}^{XY}(s) \right]^{2n} = {1 \over s^n} \sum_{j_1=0}^n \dots \sum_{j_n=0}^n & {n! \over (1!)^{j_1} \cdot ... \cdot (n!)^{j_n}} {1 \over j_1! \cdot ... \cdot j_n!} \prod_{l=0}^{k-1} (s-l) \prod_{m=1}^n \Big( \sum_{p=0}^{\lfloor m/2 \rfloor} {(m!)^2 \over 2^{3p} (m-2p)!} \sigma_X^{2p} \sigma_Y^{2p} \sigma_{XY}^{2(m-2p)}\Big)^{j_m},
\label{eq::complete.gaussian.simplified}
\end{align}
\end{widetext}
where $k=\sum_{m=1}^n j_m$. For large $s \gg n$, which is a typical case, we observe that the factor $s^{-n}$ can suppress all the summed terms except for $k = n$, where we obtain a product:
\begin{equation}
\prod_{l=0}^{n-1} (s-l) = s(s-1)...(s-n+1) \sim \mathcal{O} (s^n).
\label{eq::leading.term.simplified}
\end{equation}
In order to write the corresponding expression for the leading term, we observe that in this case $\sum_{i=1}^n j_i = \sum_{i=1}^n i j_i = n$, which is fulfilled only if $i=1$, $j_1=n$, and $j_i=0$ for $i>1$. Thus, the expression reads
\begin{widetext}
\begin{align}
\left[ F_{2n}^{XY}(s) \right]^{2n} = {1 \over s^n} {n! \over (1!)^n n!} \prod_{l=0}^{n-1} (s-l) \Big( {(1!)^2 \over 1!} \sigma_{XY}^2\Big)^n = {\sigma_{XY}^{2n} \over s^n} \prod_{l=0}^{n-1} (s-l) \sim \sigma_{XY}^{2n},
\end{align}
\end{widetext}
where we exploited the fact that $m=1$ and $\lfloor 1/2 \rfloor = 0$. The above result is apparently independent of $s$, but actually, for the fractal signals without long-term autocorrelations, it is $\sigma_{XY}^2=\sigma_{XY}^2(s) \sim s$. The fact that the signal profiles $X$ and $Y$ have been detrended by the removal of a smooth trend (typically  polynomial) does not change this behaviour since, for the sufficiently large $s$, subtracting a smooth function does not affect the fluctuations significantly. As a consequence, we obtain 
\begin{equation}
F_{2n}^{XY}(s) \sim (\sigma_{XY}^2)^{1/2} \sim s^{1/2}, 
\end{equation}
which is independent of $n$, i.e. the cross-correlations are monofractal, exactly as expected for the series of uncorrelated Gaussian random variables. This result can be generalized to $F_r^{XY}(s)$, where $r\in \mathbb{R}_{+}$, provided one notes that the generalized Hurst exponents $h(r)$ in Eq.(\ref{eq::hurst}) form a monotonic function of $r$, so $2n-1 < r < 2n$ implies that $F_{2n-1}^{XY}(s) < F_r^{XY}(s) < F_{2n}^{XY}(s)$.

That the $r$-dependence of the fluctuation functions fades out also for $r \le 0$ can in fact be seen - though on a somewhat lower level of detailed decomposition - by noticing that in the limit of large $s$ the sample covariance in a segment $\nu$ given by Eq.~(\ref{eq::covariance}) approaches the population covariance of the processes $X$ and $Y$: $f_{XY}^2(s,\nu) \approx \sigma_{XY}^2 \sim s$. By exploiting these relations, one may write:
\begin{equation}
\begin{gathered}
\left[ F_r^{XY}(s) \right]^r \approx {1 \over M_s} \sum_{\nu=0}^{M_s-1} {\rm sign} \left[ \sigma_{XY}^2 \right] |\sigma_{XY }^2|^{r/2} =\\
= {\rm sign} \left[ \sigma_{XY}^2 \right] |\sigma_{XY }^2|^{r/2} = c |\sigma^2_{XY}|^{r/2}.
\end{gathered}
\label{eq::fluctuation.function.negative}
\end{equation}
For $c=1$ it follows that
\begin{equation}
F_r^{XY}(s) \sim s^{1/2},
\label{eq::fluctuation.function.scaling}
\end{equation}
which is valid for any real $r$. If $c=-1$ for all the considered scales $s$ and if a power-law relation of the expression (\ref{eq::fluctuation.function.negative}) on $s$ is observed, we may put $\left[ F_r^{XY}(s) \right]^r \rightarrow -\left[ F_r^{XY}(s) \right]^r$ in order to avoid complex numbers~\cite{oswiecimka2014}. We then arrive at the same result (\ref{eq::fluctuation.function.scaling}). If no power-law is seen in the bivariate fluctuation function (\ref{eq::fluctuation.function.negative}), we do not deal with a fractal signal and such a case is beyond the scope of the present analysis. Of course, the special case of $Y=X$ corresponds to the standard MFDFA technique~\cite{kantelhardt2002} that deals with the correlations described by $F_r(s)$ and Eq.~(\ref{eq::fluctuation.function.negative}) involves just the variance which is always positive. 

\subsection{Heavy-tailed case}

From the Gaussian case discussed above, we now move to the case of random variables with p.d.f.s that have heavy tails, but outside the L\'evy-stable regime, and this is the most frequent case in real-world signals, including the financial ones~\cite{kwapien2012}. The structures associated with the series of uncorrelated L\'evy stable fluctuations are by now better understood and are known~\cite{nakao2000} to result in bifractals.   

Exactly thus as before, we assume that the resulting time series do not have any long-term autocorrelations. This assumption allows us to preserve the factorization of moments shown by Eq.(\ref{eq::factorization.simplified}), but now the Isserlis theorem cannot be applied and therefore the higher-order moments $\mathbb{E}(X^m Y^m)$ cannot be reduced to a combination of the second moment products. Moreover, for such distributions the higher-order moments become the larger, the heavier are the distribution tails. For the random variables with power-law tails, the moments of sufficiently high order become infinite, which forces one to reduce in principle the maximum allowed $r$ in $F_r(s)$. However, in real data there is always some cut-off of the tails, even if they behave as power-law ones for some range of random variable values, and, thus, one may consider a broader range of $r$s.

If data under consideration present heavy tails with or without cut-off, there is no longer a possibility that the factor $1/s^n$ suppresses all the terms in Eq.(\ref{eq::complete.gaussian.simplified}) except the leading one ($\sim s^n$) shown in Eq.(\ref{eq::leading.term.simplified}). The higher-order moments can become so large that the corresponding terms are not suppressed even if they are proportional to $s^{n-1}$, $s^{n-2}$ and so on. In such a case, the calculated $F_{2n}^{XY}(s)$ becomes dependent on $n$, which - for the fractal data - manifests itself via the variable generalized Hurst exponents $h(2n)$ in Eq.(\ref{eq::hurst}). In order to overcome this effect, one needs to take a sufficiently large maximum scale $s_{\rm max}$ that is able to suppress all the moments occurring with $s^{n-1}$ and the remaining terms in Eq.(\ref{eq::complete.gaussian.simplified}). This again points to $s$ as the relevant temporal scale parameter. Of course, it can be extended to sufficiently large values only when the length $T$ of entire series allows it. Once the length $T$ of the series significantly exceeds the necessary limit, its further elongation is expected not to affect much the value of $s_{\rm max}$.

It is also worth noticing at this point that on the level of MFDFA the arguments leading to Eqs.~(\ref{eq::fluctuation.function.negative}) and (\ref{eq::fluctuation.function.scaling}) can be evoked as long the variance exists. This of course is the case of heavy-tailed distributions that still do not belong to the L\'evy stable regime. Such a case is however of central interest in the present context. The L\'evy stable uncorrelated series are bifractal~\cite{nakao2000}.


\begin{figure*}
\centerline{\includegraphics[width=0.9\textwidth]{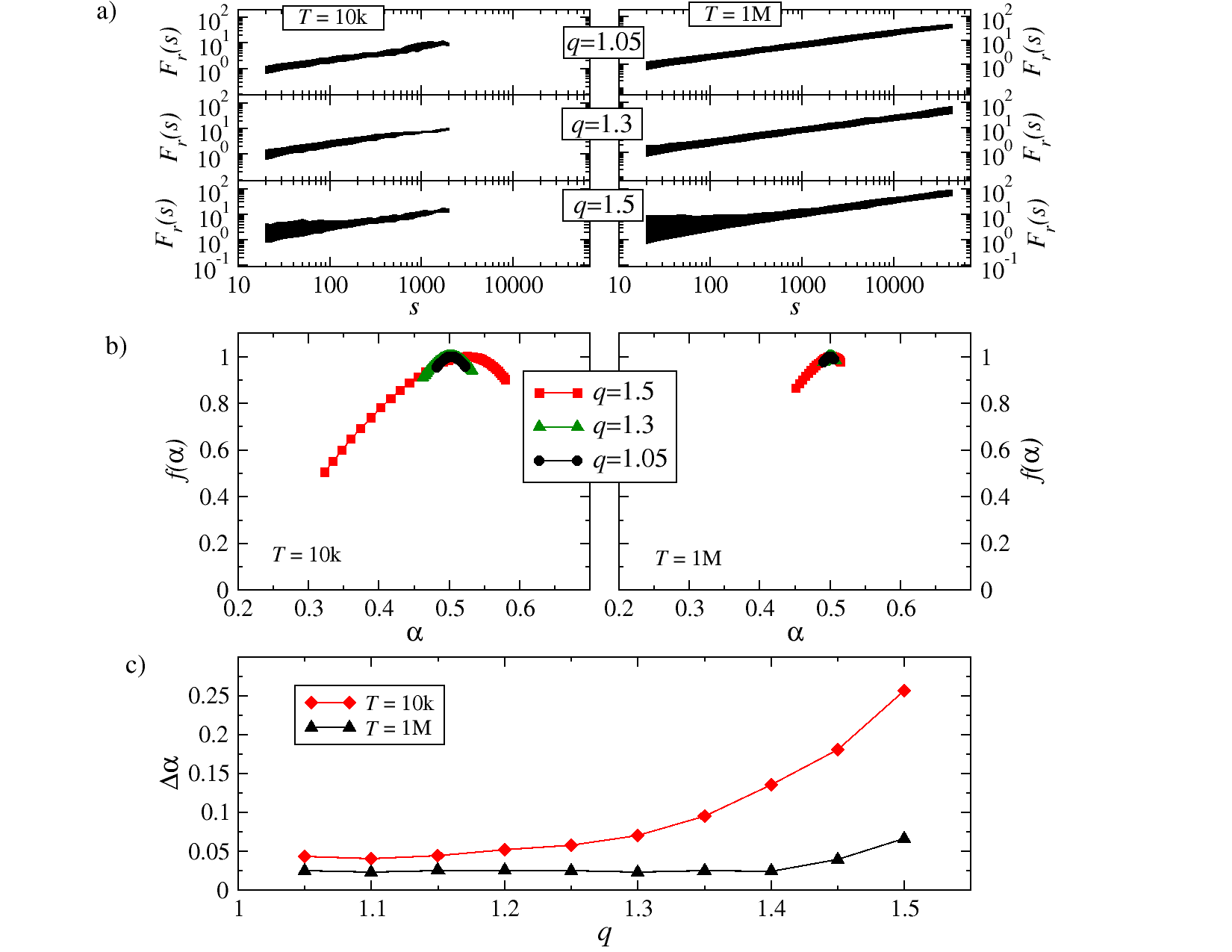}}
\caption{(a) Fluctuation functions $F_r(s)$ (MFDFA) calculated for the signals of length $T =10^4$ (left) and $T=10^6$ (right) sampled from three different $q$Gaussians. In each panel $F_r(s)$ for different moments indexed by $r \in [−4, 4]$ is displayed. (b) The corresponding singularity spectra. (c) Width $\Delta \alpha$ of these singularity spectra $f(\alpha)$ as a function of $q$ within the range $1.05 < q < 1.5$.}
\label{fig::qGaussian}
\end{figure*}

From this perspective it is clear that the $F_r^{XY}(s)$ dependence on $r$ for the heavy-tailed random variables is not a genuine multifractality as it is, for instance, in the case of time series representing the multiplicative cascades. Here it is caused solely by large values of $\mathbb{E}(X^m Y^m)$ that, in turn, are caused by the moments' weak convergence to their Gaussian counterparts expected from the central limit theorem for the non-L\'evy random variables. Because the L\'evy-stable random variables generate a bifractal (i.e., two-point) singularity spectra, a trace of this kind of behaviour due to the weak convergence is also seen in $f(\alpha)$ for the non-L\'evy random variables if the scales are relatively small. Due to the presence of noise, in real situations the bifractal $f(\alpha)$ becomes left-sided continuous and a similar effect is observed for the spectra that are not bifractal, but also left-sided due to the already-mentioned ``memory-of-bifractality'' effect. We will illustrate it with a few numerical examples in the next Section.

\section{Numerical illustrations}

Most often the multifractal characteristics of time series are expressed via the singularity spectrum $f(\alpha) = r[\alpha - h(r)] + 1$, where $\alpha=h(r) + rh'(r)$ is the singularity strength and $h(r)$ is determined by Eq.(\ref{eq::hurst}). The degree of multifractality is then measured in terms of the width $\Delta \alpha = \alpha_{\rm max} - \alpha_{\rm min}$ between the extreme values of $\alpha$ for an assumed range $r_{\rm min} \le r \le r_{\rm max}$.

In order to establish an empirical-oriented perspective on the above arguments and to see how the related effects manifest themselves in numerical experiments the four model cases are now presented. The first one (i) considers the uncorrelated time series drawn from the $q$Gaussian distribution, 
\begin{equation}
p(x) \sim e_q^{-a_q x^2} = 1 / [1+(q-1)a_q x^2]^{1/(q-1)}
\label{eq::qGaussian}
\end{equation}
in the scientific literature  proposed in connection with the concept of non-extensive entropy~\cite{tsallis2009}. This distribution appears very efficient in reproducing the fluctuations of several natural phenomena including the financial ones~\cite{rak2007}. For $q=1$ it reduces to the standard Gaussian distribution and up to $q=5/3$ it spans the whole range of heavy-tailed distributions that belong to the Gaussian basin of attraction. In particular, for $q=3/2$ it corresponds to the so-called 'inverse-cubic power-law' that is obeyed almost universally by the high-frequency return distributions in various financial markets~\cite{watorek2021a}. For the present purpose several sets of the time series of length $T=10^4$ and $T=10^6$ are generated in 100 realisations each. In Fig.~\ref{fig::qGaussian} the explicit results for $q=1.05$, $q=1.3$ and $q=1.5$ for the corresponding fluctuation functions (a) and the resulting singularity spectra (b) are shown. Clearly, as it is consistent with the above formal arguments, the effects that can erroneously be interpreted as multifractality disappear with an increasing length of the series. For the larger values of parameter $q$, as here of $q=1.5$, thus for the heavier tails in the distribution of fluctuations, this convergence to the correct, monofractal result is very slow, indeed, and the 'broom' seen at the smaller values of $s$ can confusingly be interpreted in terms of multifractality. The range of this 'broom', which can be identified as $s_{\rm max}$ of the previous section, remains largely unchanged, as anticipated. More systematically, these effects of convergence in terms of the width $\Delta \alpha$ of singularity spectrum are illustrated in the panel (c) of Fig.~\ref{fig::qGaussian}. As one can see here, an automatic (thus over the entire range of $s$ values inspected) treatment of the numerical procedure of extracting the singularity spectra from the fluctuation functions $F_r(s)$ for the uncorrelated series of $T = 10^4$ order long at the larger values of $q$ produces sizeable values of $\Delta \alpha$. In order to suppress such artifacts the length of the series needs to rich the orders of $T = 10^6$ especially for $q$ approaching 3/2. Ideally, the initial region of $s$ up to $s_{\rm max}$ should be omitted.

\begin{figure}
\centerline{\includegraphics[width=0.5\textwidth]{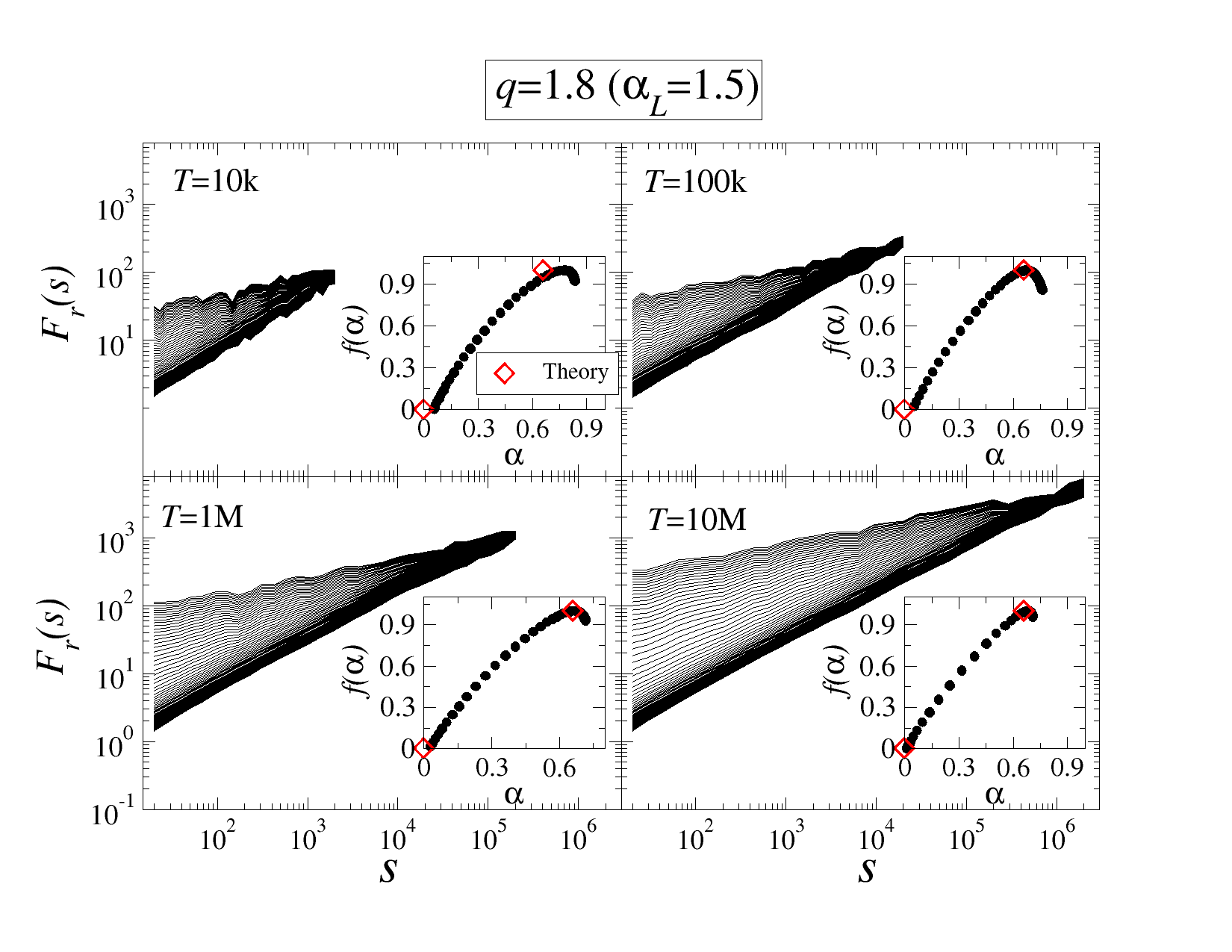}}
\caption{(Main panels) Fluctuation functions $F_r(s)$ (MFDFA) calculated for the signals sampled from a $q$Gaussian distribution with $q=1.8$, which falls into the L\'evy-stable domain. In each panel $F_r(s)$ for the signals of different length are shown from $T=10^4$ to $T=10^7$ and for $r \in [-4,4]$. (Insets) Singularity spectra $f(\alpha)$ corresponding to the fluctuation functions shown in the main panels. The theoretical values of the bifractal spectra are denoted by red diamonds.}
\label{fig::Levy}
\end{figure}

Systematically more smeared out is the situation at even larger values of $q$, those approaching the L\'evy-stable basin of attraction that extends for $5/3 < q < 3$~\cite{tsallis2006}. As it is known from analytical considerations~\cite{nakao2000}, the singularity spectrum of uncorrelated series drawn from the L\'evy-stable regime assumes a bifractal form consisting of only two points located at $(0,0)$ and $(1/{\alpha_L},1)$, where the subscript $L$ refers to the L\'evy-stable distribution. Thus, in numerical realizations, this bifractality is already felt when $q$ approaches $5/3$, which of course should not be confused with genuine multifractality. As the second (ii) numerical illustration the MFDFA realisation of a real bifractal case of $q=1.8$ which, via Eq.~(\ref{eq::qGaussian}), corresponds to $\alpha_L = 3/2$ $(\alpha_L = (3-q)/(q-1))$ is shown in Fig.~\ref{fig::Levy} for the four series of length $T = 10^4$, $T = 10^5$, $T = 10^6$ and $T = 10^7$, correspondingly. The theoretically sharp bifractal (red diamonds) is clearly smeared out and only going to really large values of $T$ allows to recognize it as bifractal from the stronger concentration of points in the MFDFA-calculated singularity spectrum $f(\alpha)$ (insets) in the vicinity of the two relevant points at $(0,0)$ and $(1/{\alpha_L,1)} $.


\begin{figure}
\centerline{\includegraphics[width=0.6\textwidth]{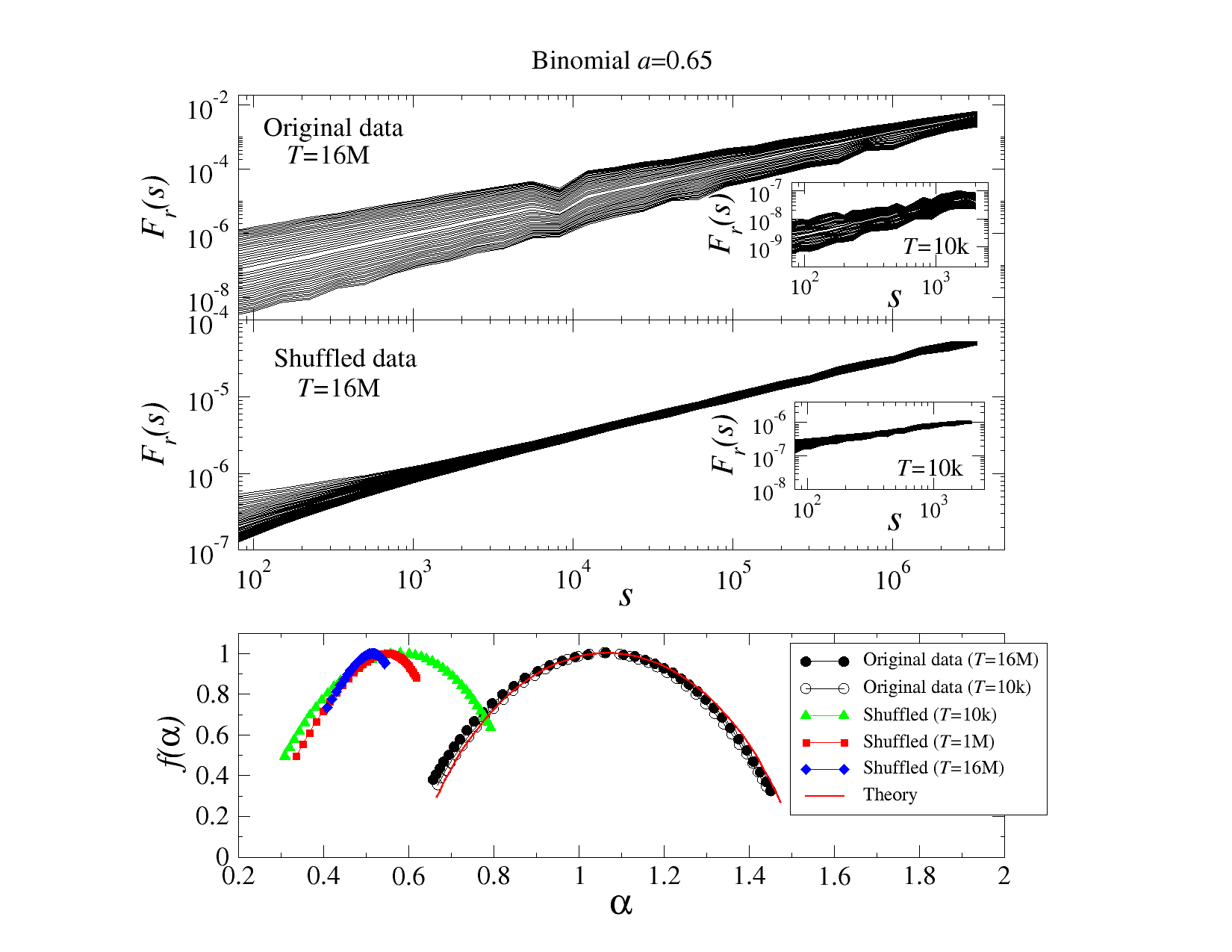}}
\caption{(Upper) Fluctuation functions $F_r(s)$ (MFDFA) within $r \in [−4, 4]$ calculated for the signals generated by the deterministic binomial cascade $a=0.65$ of length $T=16 \cdot 10^6$ (upper main panels) and for its shuffled variant (lower main panel). Analogous quantities for $T=10^4$ are shown in the corresponding insets. The resulting singularity spectra $f(\alpha)$ for the series of length $T=10^4$ and $T=16 \cdot 10^6$ correspond to the ones located on the right hand side of the lower panel while their shuffled variant is located on the left hand side of this panel. Here, in addition, the case of $T=10^6$ is shown.}
\label{fig::binomial.cascade}
\end{figure}

The third (iii) numerical example begins with a correlated series, which by construction is multifractal~\cite{kantelhardt2002}. A convenient choice is a binomial cascade. Thus a series $\{x_i\}_{i=1}^T$ of length $T=2^{k_{\rm max}}$ is generated such that, after $k_{\rm max}$ steps of iteration, $x_i=a^{n(i-1)}(1-a)^{k_{\rm max}-n(i-1)}$, where $0.5 < a < 1$ and $n(i)$ denotes the number of unities in the binary representation of $i$. For $a=0.65$, a series of length $T=2^{24}$ $(k_{\rm max})$ is generated and numerically inspected using the MFDFA algorithm. The results for different lengths $N$ are shown in Fig.~\ref{fig::binomial.cascade}. Interestingly, for the original correlated case, either short $(T=10^4)$ or long $(T=16 \times 10^6)$ series lead to essentially the same singularity spectrum. This confirms correctness of the procedure and, in fact, can be anticipated. The genuine multifractality implies self-similarity, thus a portion of the signal encodes already the generator which makes the whole. It should also be noticed that both results agree well with the theoretical predictions~\cite{kantelhardt2002}. The randomly shuffled variant of the same series, thus without any correlations, changes the picture dramatically. Clearly, the MFDFA-obtained shapes of the singularity spectra are getting narrower with their increasing length as expected based of the formal arguments presented above. The convergence to the correct result of monofractality is very slow and even for the series as long as $T=10^6$ it still develops a non-zero width.


\begin{figure}
\centerline{\includegraphics[width=0.5\textwidth]{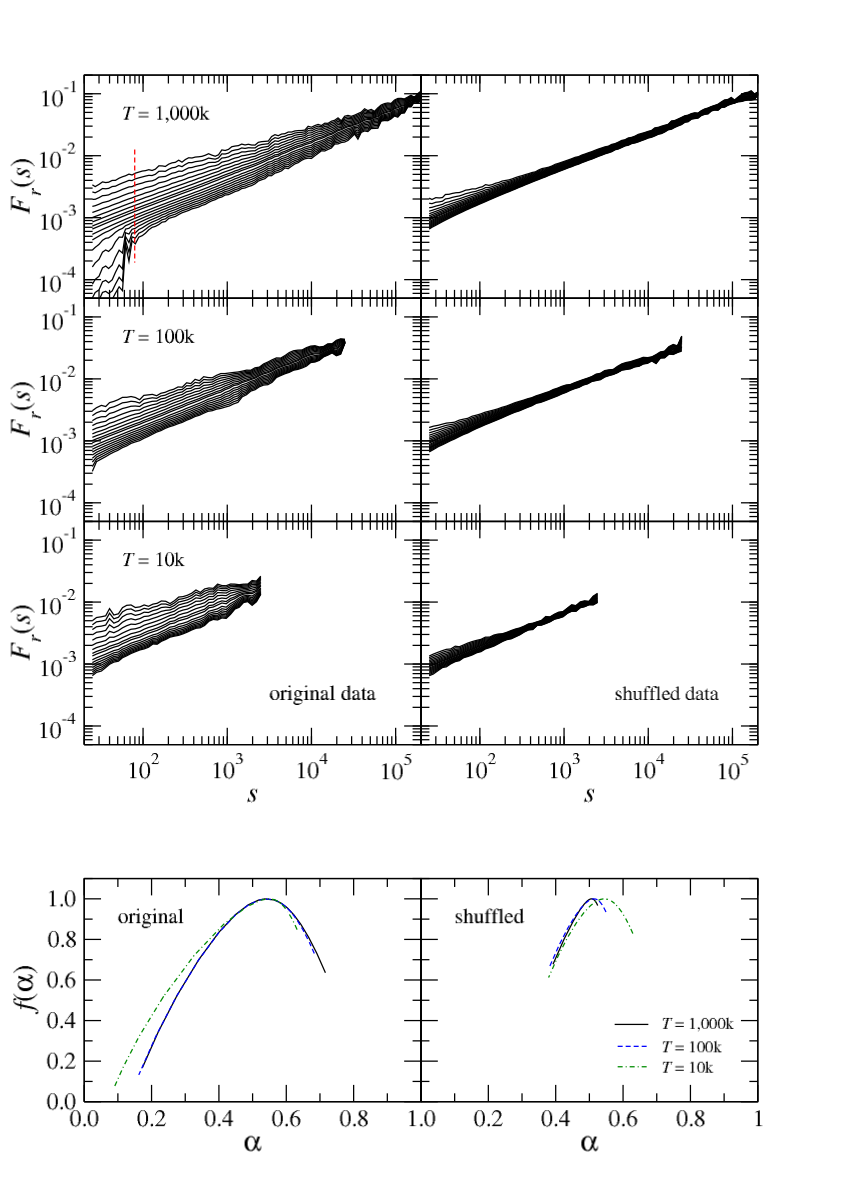}}
\caption{MFDFA multifractal characteristics based on over 1 million 1 minute bitcoin logarithmic returns expressed in USD from the period June 21, 2019 - May 15, 2021 downloaded from Kraken~\cite{kraken}, calculated for the original signal samples (left side panels) and for their randomized variants (right side panels) of length $T =10^4$, $T=10^5$ and $T=10^6$ within $r \in [−4, 4]$ moments. Fluctuation functions $F_r(s)$ are shown in the upper panels and the corresponding singularity spectra $f(\alpha)$ are displayed in the lower panels, respectively. $f(\alpha)$ are determined from the entire range of $s$. Only the strongly fluctuating part seen in $F_r(s)$ of the original signal for $T=10^6$ is cut out as indicated by the vertical dashed red line. Such a kind of fluctuations for $q<0$ takes place when several zero-valued returns appear in a row.}
\label{fig::crypto}
\end{figure}

As a final example (iv) the same multifractality-related characteristics of the real world series representing high-frequency bitcoin price changes~\cite{watorek2021a,takaishi2018} are illustrated in Fig.~\ref{fig::crypto}. This time series (available from Kraken~\cite{kraken}) represents 1-min price returns covering the period June 21, 2019 –- May 15 2021 and comprises over 1 million data points. The shape of the singularity spectrum calculated for this time series is seen to resemble the known cases from the conventional financial markets~\cite{jiang2019} and, just as for the binomial cascade studied above, the result obtained from the long series $(T \sim 10^6)$ is similar to the one obtained already for $T \sim 10^4$. Shuffling destroys the temporal correlations and then, as before, a really long series is needed to approach the monofractal limit.


\section{Summary} 

Multifractality is a concept that serves efficiently encompassing the nonlinear features of hierarchical organization in complex temporal and spatial structures. In time series the multifractal formalism quantifies the fractal dimension of the support carrying specific values of the H\"older exponents $(\alpha)$ and any such an exponent at a particular point in time series is defined relative to the neighbouring values. Therefore, it carries a potential of reflecting the nonlinear principles of the underlying hierarchical organization. Once the time series is randomized by shuffling all the related correlations are expected to get destroyed - at the first place the nonlinear ones - the neighbourhood becomes random and all the H\"older exponents are expected to approach 0.5 (the uncorrelated, trivially monofractal white noise). Of course, for a finite time series, some related dispersion of their values at around 0.5 is natural and this is what typically takes place in numerical analyses. As it is shown in the present contribution, it carries no genuine multifractality and can only be considered as an apparent effect of the short scales. To eliminate such an effect, the length of time series needs to be sufficiently large, approaching even millions of data points when the pdf tails are significantly thicker than the Gaussian ones. In this connection a more precise criterion needs to be formulated than the one in ref.~\cite{kantelhardt2002} where it is stated that the series of uncorrelated power-law distributed values has rather bifractal properties. The present study clarifies it further that it is bifractal, indeed, but only in the L\'evy stable regime of the distribution. The power-law distributed uncorrelated fluctuations with the tails thinner such that they are L\'evy unstable, thus belong to the Gaussian basin of attraction, lead to the monofractal dynamics. This latter case comprises a broad range of real-world phenomena including in particular the financial ones quite universally~\cite{watorek2021b} obeying the so-called inverse cubic power-law~\cite{gopikrishnan1998}. Even more so this last case of monofractality applies to the stretched exponential distributions appearing quite frequently in inter-transaction time fluctuations~\cite{kwapien2022} and in many other natural phenomena~\cite{laherrere1998}. Numerically, similar effects are observed when the wavelet based algorithms are used (see e.g. Fig.~2 in  ref.~\cite{drozdz2009}). On the other hand, once present, the global features of genuine multifractality are encoded already in a relatively short portion of the whole signal as the above numerical experiments document it. This may reflect another manifestation of the conjectured $q$-generalized Central Limit Theorem~\cite{moyano2006} which allows a broader class of stable distributions when the correlations are present.

\end{document}